\documentclass[twocolumn,showpacs,amsmath,amssymb,pra,longbibliography]{revtex4-1}

\usepackage[pdftex]{graphicx} % Include figure files
\usepackage{dcolumn}  % Align table columns on decimal point
\usepackage{bm}       % bold math
\usepackage[usenames,dvipsnames]{color}
\definecolor{URLCOL}{rgb}{0,0.52,0.83} %external link color
\definecolor{LINKCOL}{rgb}{0.05,0.5,0} %internal link color
\definecolor{CITECOL}{rgb}{0.25,0,0.48} %link to bibliography
\usepackage{epstopdf}

\usepackage[pdftex,bookmarks,breaklinks,bookmarksopen,bookmarksnumbered,colorlinks,linkcolor=LINKCOL,linktocpage,citecolor=CITECOL,urlcolor=URLCOL,pdfpagemode=UseOutline,pdftex]{hyperref}

\usepackage[normalem]{ulem}
\usepackage{booktabs}
\usepackage{comment}
\usepackage{natbib}
\definecolor{TITLECOL}{rgb}{0.1,0.2,0.7} %title color
\definecolor{SECOL}{rgb}{0.1,0.2,0.7} %sec color
\definecolor{CONTENTSCOL}{rgb}{0.1,0.2,0.7} %can choose the table of contents title to have same color as sec
\definecolor{SSECOL}{rgb}{0.25,0,0.48} %ssection color
\definecolor{SSSECOL}{rgb}{0.2,0.08,0.53} %subsubsection color  0.2,0.08,0.53

\def\coloredtitle#1{\title{\textcolor{TITLECOL}{#1}}} %title color
\def\coloredauthor#1{\author{\textcolor{CITECOL}{#1}}} %author color

\definecolor{URLCOL}{rgb}{0,0.17,0.43} %external link color
\definecolor{LINKCOL}{rgb}{0.05,0.4,0} %internal link color
\definecolor{CITECOL}{rgb}{0.35,0,0.48} %link to bibliography

\def\sss{\scriptscriptstyle\rm}

\def\bea{\begin{eqnarray}}
\def\eea{\end{eqnarray}}
\def\ben{\begin{equation}}
\def\een{\end{equation}}
\def\benu{\begin{enumerate}}
\def\enu{\end{enumerate}}

\def\bei{\begin{itemize}}
\def\eei{\end{itemize}}
\def\beit{\begin{itemize}}
\def\eit{\end{itemize}}
\def\benu{\begin{enumerate}}
\def\enu{\end{enumerate}}

\def\half{\frac{1}{2}}

\def\x{_{\sss X}}
\def\c{_{\sss C}}
\def\s{_{\sss S}}
\def\F{_{\sss F}}
\def\xc{_{\sss XC}}

\def\ee{_{\rm ee}}

\def\LDA{^{\rm LDA}}

\def\LSD{^{\rm LSD}}
\def\unif{^{\rm unif}}
\def\up{_\uparrow}
\def\dn{_\downarrow}
\def\HF{^{\rm HF}}
\def\QC{^{\rm QC}}

\def\n{n}

\def\Tabref#1{Table \ref{#1}}

\def\Eqref#1{Eq.\ \eqref{#1}}
\def\Secref#1{Section \ref{#1}}
\def\Figref#1{Fig.\ \ref{#1}}
\def\Ref#1{Ref.\ \cite{#1}}
\def\Refs#1{Refs.\ \cite{#1}}

\def\sec#1{\section{\textcolor{SECOL}{#1}}}
\def\ssec#1{\subsection{\textcolor{SSECOL}{#1}}}

\def\e{_\text{e}}

\begin{document}

\coloredtitle{
Reference electronic structure calculations in one dimension
}
\coloredauthor{Lucas O.\ Wagner}
\affiliation{Department of Physics and Astronomy, University of California, Irvine, CA 92697}
\coloredauthor{E.M. Stoudenmire}
\affiliation{Department of Physics and Astronomy, University of California, Irvine, CA 92697}
\coloredauthor{Kieron Burke}
\altaffiliation[Also at ]{Department of Chemistry, University of California, Irvine, CA 92697}
\affiliation{Department of Physics and Astronomy, University of California, Irvine, CA 92697}
\coloredauthor{Steven R.\ White}
\affiliation{Department of Physics and Astronomy, University of California, Irvine, CA 92697}
\date{\today}

\begin{abstract}
Large strongly correlated systems provide a challenge to modern electronic structure
methods, because standard density functionals usually fail and traditional quantum
chemical approaches are too demanding.  The density-matrix renormalization group
method, an extremely powerful tool for solving such systems, has recently been extended
to handle long-range interactions on real-space grids, but is most efficient in one dimension
where it can provide essentially arbitrary accuracy.   Such 1d systems therefore
provide a theoretical laboratory for studying strong correlation and developing density
functional approximations to handle strong correlation, {\em if} they mimic three-dimensional
reality sufficiently closely.   We demonstrate that this is the case, and provide reference
data for exact and standard approximate methods, for future use in this area.
\end{abstract}

\pacs{71.15.Mb, %DFT for condensed matter
31.15.E-, %DFT for atomic systems
05.10.Cc %renormalization group methods
}

\maketitle
%\coltableofcontents
%\sf

%%%%%%%%%%%%%%%%%%%%%%%%%%%%%%%%%%%%%%%%%%%%%%%%%%%%%%%%%%%%%%%%%%%%%%%%%%%%%%%%%%%%%%%%%%%%%%%%%%%%%%%%%%%%%%%%%%%%%%%%
%%%%%%%%%%%%%%%%%%%%%%%%%%%%%%%%%%%%%%%%%%%%%%%%%%%%%%%%%%%%%%%%%%%%%%%%%%%%%%%%%%%%%%%%%%%%%%%%%%%%%%%%%%%%%%%%%%%%%%%%
%%%%%%%%%%%%%%%%%%%%%%%%%%%%%%%%%%%%%%%%%%%%%%%%%%%%%%%%%%%%%%%%%%%%%%%%%%%%%%%%%%%%%%%%%%%%%%%%%%%%%%%%%%%%%%%%%%%%%%%%

\sec{Introduction and philosophy}

Electronic structure methods such as density functional theory (DFT) are excellent tools for investigating the properties
of solids and molecules---except when they are not.  Standard density functional approximations in the Kohn--Sham (KS)
framework \cite{KS65} work well in the weakly correlated regime \cite{CCSA98,HN00,CZFF11}, but these same approximations can fail miserably
when the electrons become strongly correlated \cite{CMY08}.
A burning issue in practical materials science today is the desire to develop
approximate density functionals that work well, even for strong correlation.
This has been emphasized in the work of Cohen {\em et al.}~\cite{CMY08,MCY09}, where even the
simplest molecules, H$_2$ and H$_2^+$, exhibit features essential
to strong correlation when stretched.

Many approximate methods, both within and beyond DFT, are currently being developed for
tackling these problems, such as the HSE06 functional \cite{HSE06} or the 
dynamical mean-field theory \cite{GKKR96}.
Their efficacy is usually judged by comparison with experiment over a range of
materials, especially in calculating gaps and predicting correct magnetic phases.
But such comparisons are statistical and often mired in controversy,
due to the complexity of extended systems.

In molecular systems, there is now a large variety of traditional (ab initio) methods
for solving the Schr\"odinger equation with high accuracy, so approximate methods
can be benchmarked against highly-accurate results,
at least for small molecules \cite{J06}.   Most such methods have not yet been reliably adopted
for extended systems, where quantum Monte Carlo (QMC) \cite{NU99} has become one of the few ways
to provide theoretical benchmarks \cite{Mc04}.  But QMC is largely limited to the ground state
and is still relatively expensive.
Much more powerful and efficient is the 
density-matrix renormalization group (DMRG) \cite{White:1992,White:1993a,Schollwock:2005},
which has scored some impressive successes in extended systems \cite{CS11}, but whose
efficiency is greatest in one-dimensional systems.

A possible way forward is therefore to study simpler systems, defined only in
one dimension, as a theoretical laboratory for understanding strong correlation.
In fact, there is a long history of doing just this, but using lattice Hamiltonians
such as the Hubbard model \cite{H63}.  While such methods do yield insight into strong
correlation, such lattice models differ too strongly from real-space models
to learn much that can be directly applied to DFT of real systems.  However,
DMRG has recently been extended to treat long-range interactions
in real space \cite{SWWB11}.  This then begs the question:  Are one-dimensional analogs sufficiently
similar to their three-dimensional counterparts to allow us to learn anything
about real DFT for real systems?

In this paper, we show that the answer is definitively yes by carefully and
precisely calculating many exact and approximate properties of small systems.
We use DMRG for the exact calculations and the one-dimensional local-density
approximation for the DFT calculations \cite{HFCV11}.   In passing, we establish many
precise reference values for future calculations.  Of course,
the exact calculations could be performed with any traditional method
for such small systems, but DMRG is ideally suited to this problem, and
will in the future be used to handle 1d systems too correlated for even the
gold-standard of ab initio quantum chemistry, CCSD(T).

Thus our purpose here is not to understand real chemistry, which is intrinsically
three dimensional, but rather to check that our 1d theoretical laboratory is
qualitatively close enough to teach us lessons about handling strong correlation
with electronic structure theories, especially density functional theory.

%-----------------------------
\begin{figure}[t]
\includegraphics[width=\columnwidth]{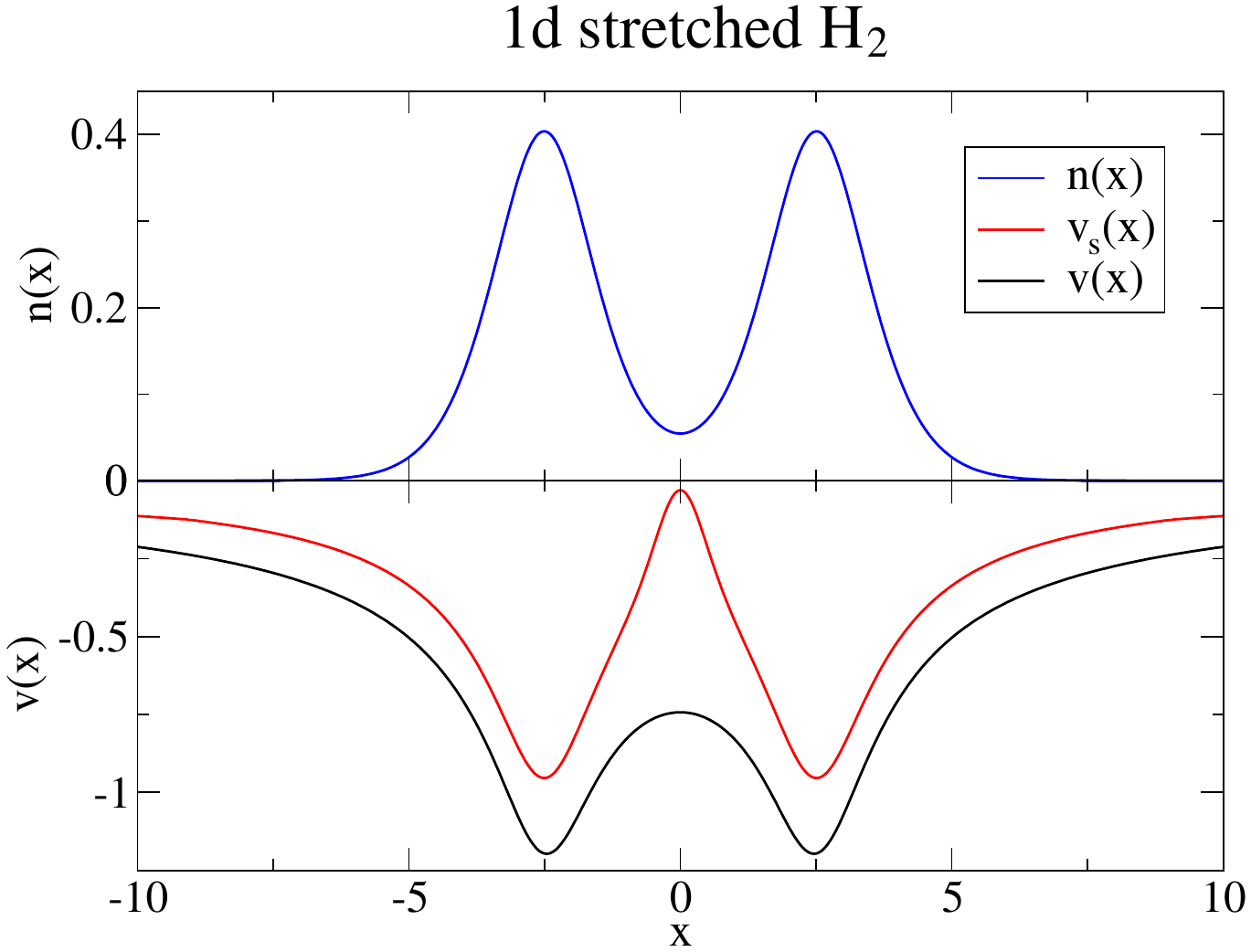}
\caption{
The KS potential for a stretched hydrogen molecule found from interacting electrons in 1d.
\label{fig:H2}
}
\end{figure}
%-----------------------------

Our results are illustrated in \Figref{fig:H2}, which shows 1d H$_2$ with soft-Coulomb interactions,
plotted in atomic units.
The exact density was found by DMRG and inverted to find the corresponding
exact KS potential, $v\s(x)$.  The bond has been stretched beyond the Coulson--Fischer point,
where Hartree--Fock and DFT approximations do poorly, as discussed further in \Secref{s:cor}.  
We comment here that a
strong XC contribution to the KS potential is needed to reproduce the exact density in the
bond region \cite{HTR09}.
Calculations to obtain the KS potential have often been performed for few-electron systems in
3d in the past \cite{UG94,PNW03}, but our method allows exact treatment of systems with many electrons.
In another paper \cite{SWWB11}, we show how powerful our DMRG method is, by solving a chain of 100 
1d H atoms.  All such calculations were
previously unthinkable for systems of this size, and unreachable by any other method.
We have applied these techniques to perform the first ever Kohn--Sham 
calculations using the exact XC functional, essentially implementing the exact Levy--Lieb
constrained search definition of the functional, which we will present in yet another paper.

\sec{Background in DMRG}\label{s:bkgnd}

The density matrix renormalization group (DMRG) is a powerful numerical method for
computing essentially exact many-body ground-state wavefunctions \cite{White:1992,White:1993a}.
Traditionally, DMRG has been applied to 1d and quasi-2d finite-range lattice models for strongly correlated electrons \cite{Schollwock:2005}.
DMRG has also been applied to systems in quantum chemistry, where the long-range Coulomb interaction is distinctive.
The Hamiltonians which have been studied in this context include the Pariser--Parr--Pople model \cite{FOZ98} 
and the second-quantized form of the Hartree--Fock equations, where  lattice sites represent electronic orbitals \cite{White:1999,Chan:2008,CS11}.

DMRG works by truncating the exponentially large basis of the full Hilbert space down to a much smaller
one which is nevertheless able to represent the ground-state wavefunction accurately.
Such a truncation would be highly inefficient in a real-space, momentum-space, or orbital basis; 
rather, the most efficient basis consists of the eigenstates of the reduced density matrix computed across bipartitions  
of the system \cite{White:1992}.
A DMRG calculation proceeds back and forth through a 1d system in a sweeping pattern, first optimizing the
ground-state in the current basis then computing an improved basis for the next step.
By increasing the number of basis states $m$ that are kept, DMRG can find the wavefunction to arbitrary accuracy.

The computational cost of DMRG scales as \mbox{$N_s m^3$} where $N_s$ is the number of lattice sites. 
For gapped systems in 1d, the number of states $m$ required to compute the ground-state to a specified accuracy
is independent of system size, allowing DMRG to scale linearly with $N_s$.
For gapless or critical systems, the $m$ needed grows logarithmically with system size, making the scaling only slightly worse.
The systems considered here have a relatively low total number of electrons such that the number of states $m$ required
is small, often less than $100$. This in turn enables us to work with the very large numbers of sites ($N_s\sim 1000-5000$) 
needed to reach the continuum, as described in more detail below.

\sec{Methodology}

%%%%%%%%%%%%%%%%%%%%%%%%%%%%%%%%%%%%%%%%%%%%%%%%%%%%%%%%%%%%%%%%%%%%%%%%%%%%%%%%%%%%%%%%%%%%%%%%%%%%%%%%%%%%%%%%%%%%%%%%

To apply DFT in its natural context---in the continuum---we shall consider a model
of soft-Coulomb interacting matter \cite{ESJ89,TGK08,CFD11}, where the electron repulsion has the form
\ben
v\ee(u) = \dfrac{1}{\sqrt{u^2 + 1}}, \label{vee}
\een
and the interaction between an electron and a nucleus with charge $Z$ and location $X$ is
\ben
v(x) = -Z v\ee(x - X).
\een
The soft-Coulomb interaction is chosen to avoid divergences when
particles are close to one another, and has been used to study molecules in intense laser fields \cite{ESJ89,TGK08}.
The wavefunctions and densities within this model
lack the cusps present in 3d Coulomb systems.  However, the challenge presented
by the long-range interactions in 3d Coulomb systems remains for these 1d model systems.

%%%%%%%%%%%%%%%%%%%%%%%%%%%%%%%%%%%%%%%%%%%%%%%%%%%%%%%%%%%%%%%%%%%%%%%%%%%%%%%%%%%%%%%%%%%%%%%%%%%%%%%%%%%%%%%%%%%%%%%%

%%%%%%%%%%%%%%%%%%%%%%%%%%%%%%%%%%%%%%%%%%%%%%%%%%%%%%%%%%%%%%%%%%%%%%%%%%%%%%%%%%%%%%%%%%%%%%%%%%%%%%%%%%%%%%%%%%%%%%%%
%%%%%%%%%%%%%%%%%%%%%%%%%%%%%%%%%%%%%%%%%%%%%%%%%%%%%%%%%%%%%%%%%%%%%%%%%%%%%%%%%%%%%%%%%%%%%%%%%%%%%%%%%%%%%%%%%%%%%%%%
%%%%%%%%%%%%%%%%%%%%%%%%%%%%%%%%%%%%%%%%%%%%%%%%%%%%%%%%%%%%%%%%%%%%%%%%%%%%%%%%%%%%%%%%%%%%%%%%%%%%%%%%%%%%%%%%%%%%%%%%

Although many methods could be used to solve these 1d systems,
DMRG allows us to work efficiently with any arbitrary 1d real-space system,
without the need to develop a basis for every 1d element. 
We enable DMRG to operate in the continuum by discretizing over a fine real-space grid. 
With a lattice spacing of $a$, the real-space Hamiltonian for a 1d system
becomes in second quantized notation,
\begin{align}
H & = \sum_{j,\sigma}  \frac{-1\ \ }{2 a^2} (c^\dagger_{j\sigma} c_{j+1,\sigma} + c^\dagger_{j+1,\sigma} c_{j\sigma}) -\tilde\mu\, n_{j\sigma} \nonumber \\
& \mbox{} + \sum_j v^j \, n_j + \half \sum_{i j} v\ee^{ij}\ n_i \, (n_j-\delta_{ij}),    \label{eqn:discrete_H}
\end{align}
where \mbox{$\tilde\mu = \mu-1/a^2$}, \mbox{$v^j = v(j\, a)$} and \mbox{$v\ee^{ij} = v\ee(|i-j|\, a)$}.
The $\delta_{ij}$ in the last term cancels self interactions.
The operator $c^\dagger_{j\sigma}$ creates (and $c_{j\sigma}$ annihilates) an electron of spin $\sigma$ on site $j$,
$n_j = n_{j \uparrow} + n_{j \downarrow}$, and $n_{j\sigma} = c^\dagger_{j\sigma} c_{j\sigma}$.  The hopping terms $c^\dagger_{j\sigma} c_{j+1,\sigma}$ 
(and complex conjugate) come about from a finite-difference approximation to the second derivative.
Like the second-quantized Hamiltonians considered in quantum chemistry, this Hamiltonian corresponds to
an extended Hubbard model;
\Eqref{eqn:discrete_H}, however, is motivated from a desire to study the 1d continuum alongside familiar DFT approximations.
Because we require that the potentials and interactions vary slowly on the scale of the grid spacing,
the low-energy eigenstates of the discrete Hamiltonian \eqref{eqn:discrete_H} will approximate the continuum system
 to very high accuracy.  Moreover, we check convergence with respect to lattice spacing.
Because our potentials---and thus our ground-state densities---vary slowly on the scale of the grid spacing, 
we can accelerate convergence by using a higher-order finite-difference approximation
to the kinetic energy operator; this simply amounts to including more hopping terms in \Eqref{eqn:discrete_H}.

Even in its discretized form the Hamiltonian Eq.~(\ref{eqn:discrete_H}) represents a challenge 
for DMRG because of the long-range interactions. 
Including all $N_s\!^2$ interaction terms, where $N_s$ is the number of lattice sites,
would make the calculation time scale as $N_s\!^3$ overall. 
Fortunately, an elegant solution has been recently developed \cite{Pirvu:2010} which involves
rewriting the Hamiltonian as a matrix product operator (MPO)---a string of operator-valued
matrices. 
This form of the Hamiltonian is very convenient for DMRG, and MPOs naturally encode 
exponentially-decaying long-range interactions \cite{McCulloch:2007}.
Assuming that our interaction $v\ee(u)$ can be approximated by a sum of exponentials, 
the the calculation time scales only linearly with the number of exponents $N_\text{exp}$ used.
This reduces the computational cost from $N_s\!^3$ to $N_s \, N_\text{exp}$.
In practice, for our soft-Coulomb interactions and modest system sizes ($N_s < 1000$), 
we find that only $N_\text{exp} = 20$ exponentials are needed to 
obtain an accuracy of $10^{-5}$ in our approximate $v\ee(u)$.  The largest $N_\text{exp}$ we use in this
paper is $60$, 
which is necessary to find the equilibrium bond length of 1d H$_2$ accurate to $\pm 0.01$ bohr (a system with $N_s \approx 2000$).

For technical reasons, we take all of our systems to have open (or box) boundary conditions. 
This has no adverse effect on our results because we can extend the grid well past our edge atoms.
The extra grid sites cost almost no extra simulation time due to the very low density of electrons in the edge regions.
To evaluate the dependence of the energy on these edge effects and the
grid size, consider \Tabref{t:basis}.  This table shows the convergence
of the 1d model hydrogen atom ground-state energy with respect to the lattice spacing $a$
and the distance $c$ from the atom to the edge of the system, using
the second-order finite difference approximation for the kinetic energy, as in
\Eqref{eqn:discrete_H}.  Our best estimate for the 1d H atom energy is \mbox{-0.66977714}, converged to 
at least microhartree accuracy, which differs slightly from that of Eberly {et al.}, who were
the first to consider the soft-Coulomb atom and its eigenstates \cite{ESJ89}.

\begin{table}
%\raggedright
%\begin{tabular}{|c|rrrr|}
\begin{tabular*}{\columnwidth}{@{\extracolsep{\fill}}crrrr}
\hline
$a$ 	&~~$c= 8$	&~~$c = 9$	 &~$c = 10$&~$c\rightarrow \infty$	\\
\hline
~0.1000~ &	-81.50	&	-82.30	&	-82.40	&	-82.41	\\
0.0500	&	-19.58	&	-20.46	&	-20.57	&	-20.58	\\
0.0200	&	-2.22	&	-3.16	&	-3.27	&	-3.29	\\
0.0100	&	0.27	&	-0.68	&	-0.80	&	-0.82	\\
0.0050	&	0.90	&	-0.07	&	-0.18	&	-0.20	\\
0.0025	&	1.06	&	0.09	&	-0.03	&	-0.05	\\
$\rightarrow 0$	&1.12	&	0.14	&	0.02	&	0.00	\\
\hline

\end{tabular*}
\caption{ Convergence of model hydrogen energy with respect to lattice spacing $a$ and 
distance $c$ from the atom to the edge of the system, with differences in units
of microhartree from the 
infinite continuum extrapolation of $E = -0.66977714$.
\label{t:basis}
}
\end{table}

In addition to the accurate many-body solutions offered by DMRG, we can also look at
approximate solutions given by standard quantum chemistry tools.
Hartree--Fock (HF) theory can be formulated for these 1d systems by trivially changing
out the Coulomb interaction for the soft-Coulomb.  The exchange energy is then:
\bea
E\x &=& -\half \sum_\sigma \sum_{i,j=1}^{N_\sigma} \int dx \int dx' \,v\ee(x-x') \times \nonumber \\
&&\quad \quad \phi_{i\sigma}(x)\phi_{j\sigma}(x) \phi_{j\sigma}(x') \phi_{i\sigma}(x').\label{Ex}
\eea
In performing HF calculations, instead of using an orbital basis
of Gaussians or some other set of functions, our ``basis set'' will be the grid,
as in \Eqref{eqn:discrete_H}.  This 
simple and brute force approach allows us a great degree of flexibility, 
but is only computationally tractable in 1d.

\begin{figure}[h]
\includegraphics[width=\columnwidth]{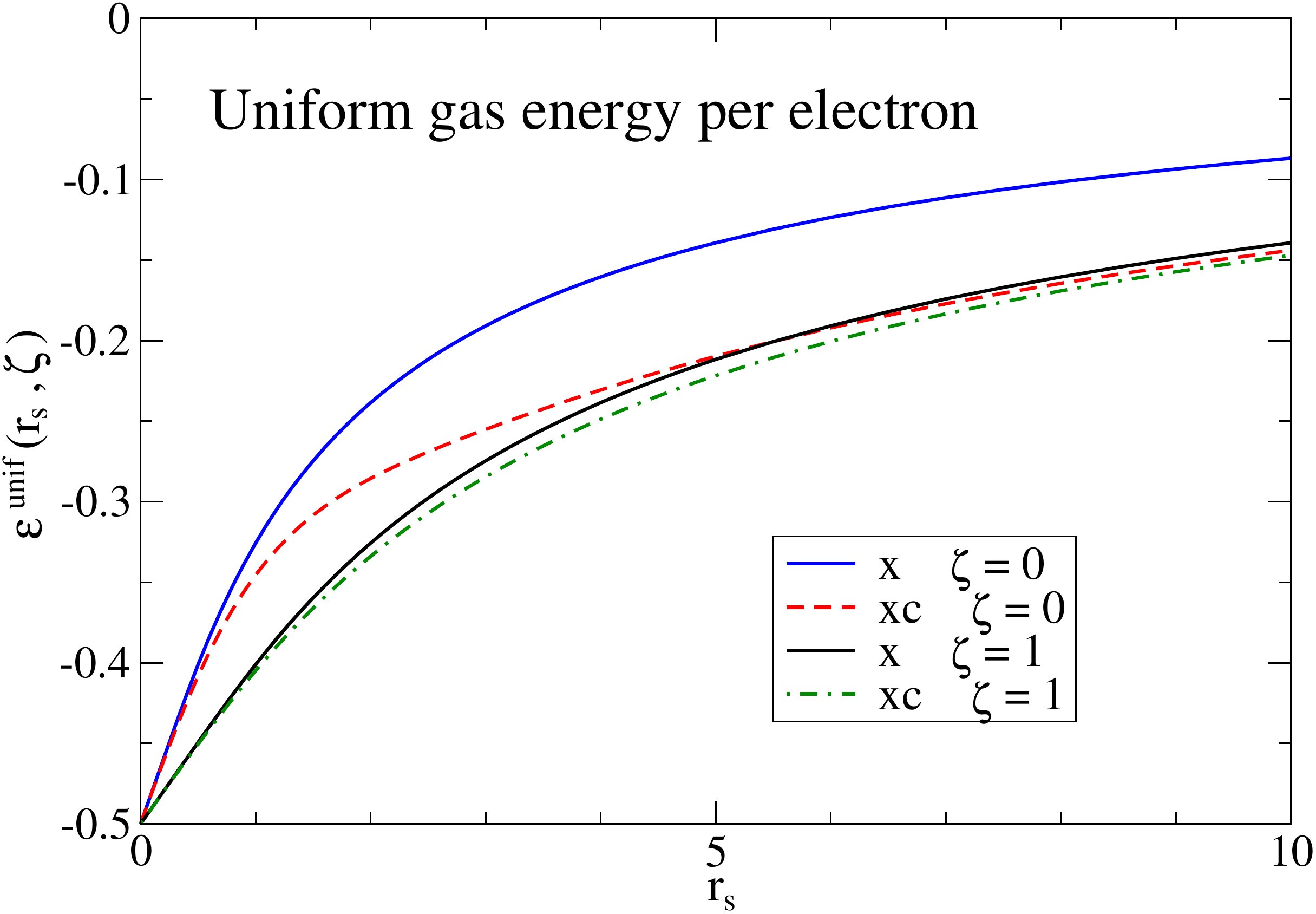}
\caption{
Parametrization of the LDA exchange and exchange-correlation energy densities per electron for
polarized $\zeta = 1$ and unpolarized $\zeta = 0$ densities \cite{HFCV11}.
\label{fig:LDAexc}
}
\end{figure}

In this setting we also implement DFT.
As mentioned in the introduction, DFT has been applied directly to lattice models.
But our model and interaction are meant to mimic the usual application of DFT to the continuum.
In particular, the LDA functionals we will use are similar to their 3d counterparts,
unlike the Bethe ansatz LDA (BALDA), which has a gap built in \cite{LSOC03,FVC11}.
One calculates
the LDA exchange energy by taking the exchange energy density per electron for a uniform gas
of density $n$,  namely $\epsilon^\text{unif}\x(n)$,
and then integrating it along with the electronic density:
\ben
E\x\LDA[\n]= \int dx\, n(x)\, \epsilon^\text{unif}\x\big( n(x) \big).
\een
We find $\epsilon^\text{unif}\x(n)$ by evaluating \Eqref{Ex} with the KS orbitals of a uniform gas.
For a uniform gas, the KS
orbitals are the eigenfunctions of a particle in a box, whose edges
are pushed to infinity while the bulk density is kept fixed.
Because the interaction has a length-scale, i.e.\ $v\ee(\gamma u) \neq \gamma^p v\ee(u)$
for some $p$, even exchange is not a simple function.
One finds:
\ben
\epsilon\x\unif(n) = -n\, f(k\F)/2,
\een
where $k\F=\pi n/2$ is the Fermi wavevector and 
\ben
f(z) = \int_0^\infty dy\, \frac{\sin^2y}{y^2} \frac{1}{{\sqrt{z^2+y^2}}}.
\een
In fact, $f$ is  
related to the Meijer $G$ function:\footnote{Information
about the Meijer $G$ function can be found online at \href{http://mathworld.wolfram.com/MeijerG-Function.html}
{http://mathworld.wolfram.com/MeijerG-Function.html}.}
\ben
f(z)=
G^{2,2}_{2,4}
\left(   \begin{array}{c}\frac{1}{2}, 1 \\ \frac{1}{2}, \frac{1}{2}, -\frac{1}{2}, 0 \end{array} \Big|
z^2 \right)/(4 z)\,.
\een
We write $r_s=1/(2n)$ as the average spacing between electrons in 1d.
In \Figref{fig:LDAexc}, we show the exchange energy per electron for the unpolarized
gas as a function of $r_s$.  For small $r_s$ (high density), 
$\epsilon\x^\text{unif}\rightarrow -1/2 + 0.203\,r_s$; for large $r_s$ (low density),
$\epsilon\x^\text{unif}\rightarrow -0.291/r_s -\ln(r_s)/(4r_s)$.
For contrast, in 3d, the exchange  energy per electron is always $-0.458/r_s$ \cite{D30},
where $r_s = (3/(4\pi n))^{1/3}$.

In practice, we do not use pure DFT, but rather spin-DFT, in which all quantities
are considered functionals of the up and down spin densities.  In that case, we need
LSD, the local spin-density approximation.  For exchange, there is a simple spin
scaling relation that tells us \cite{FNM03}
\ben
\epsilon\x\unif (\n\up,\n\dn) = - \n\, \sum_{\sigma=\pm 1}
(1+\sigma\zeta)^2\, f(k\F (1+\sigma\zeta))/4,
\een 
where $\zeta = (n_\uparrow - n_\downarrow)/n$ is the polarization. 
This is less trivial than for simple Coulomb repulsion.  At high densities, there
is no increase in exchange energy due to spin polarization, while there is a huge
increase (tending to a factor of 2) at low density, as shown by the solid black line in \Figref{fig:LDAexc}.
In fact, $\epsilon\x\unif(r_s,\zeta=1)=\epsilon\x\unif(r_s/2,\zeta=0)$.

\begin{figure}[t]
\includegraphics[width=\columnwidth]{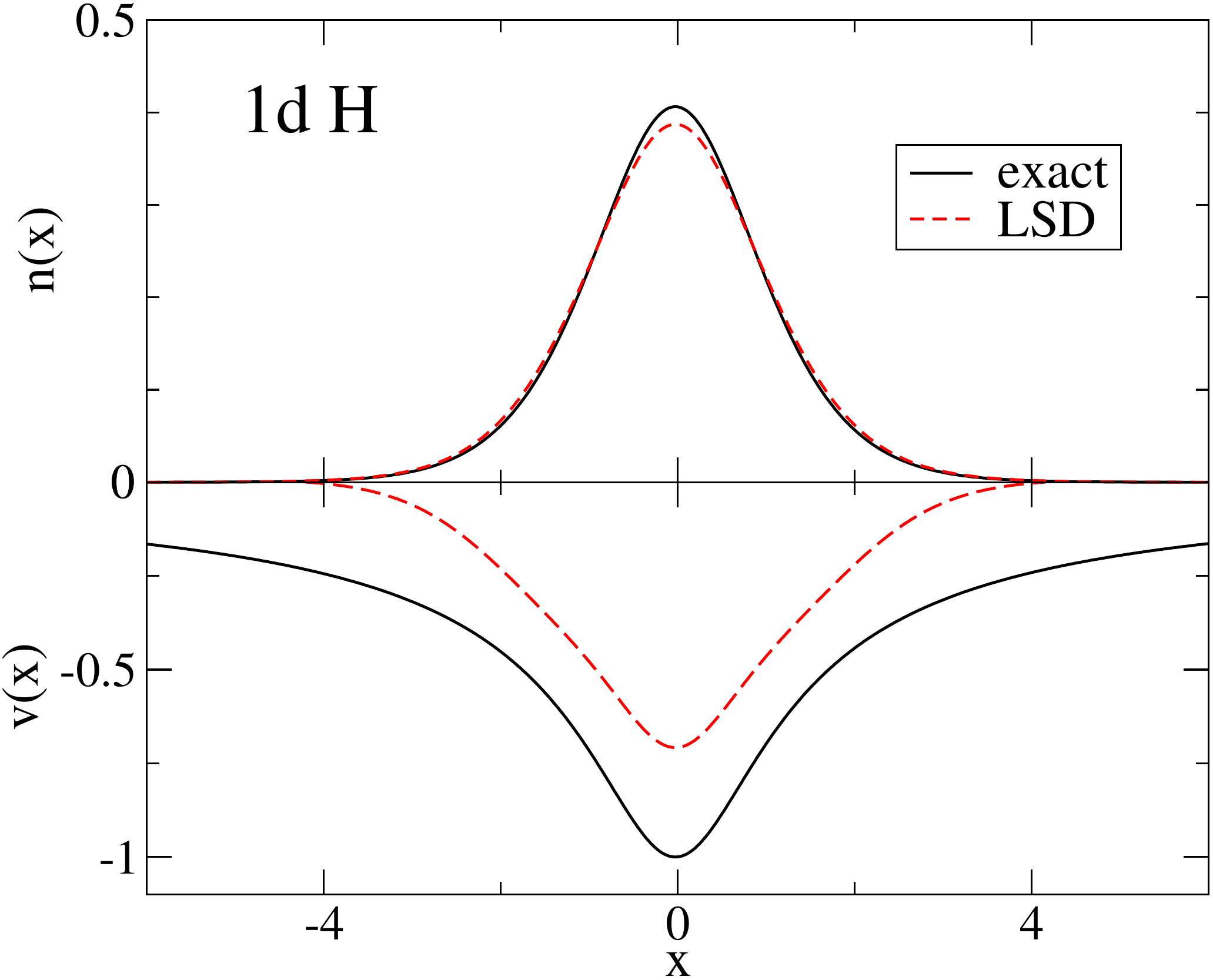} %with KS lda potential
\caption{
The hydrogen atom with both exact and LSD densities, as well as
the LSD KS potential.
\label{fig:H}
}
\end{figure}

To complete LDA, we need the correlation energy density of the uniform gas at various
densities and polarizations.
We are very fortunate to be able to make use of the
pioneering work of \Ref{HFCV11}, which performs just such a QMC calculation
and parametrizes the results, yielding accurate values for 
$\epsilon\c^\text{unif}(r_s,\zeta)$, which are also plotted for the unpolarized
and fully polarized cases in \Figref{fig:LDAexc}.
These curves are not qualitatively similar to the 3d $\epsilon\xc^\text{unif}(r_s,\zeta)$.  
For these 1d model systems,  the fully polarized electrons almost completely
avoid one another at the exchange level, so that correlation barely decreases their energy
for any value of $r_s$.  For unpolarized electrons, the effect of correlation is to make
them avoid each other entirely for low densities ($r_s > 5$)
and the XC energy per electron becomes independent of polarization.
However, for unpolarized electrons at high density, correlation vanishes with $r_s$, and 
exchange dominates, as in the usual 3d case.  For moderate $r_s$ values, the correlation
contribution grows with $r_s$, as shown by the red dashed line of \Figref{fig:LDAexc}.
To give an idea of what range of $r_s$ is important, for the hydrogen atom of 
\Figref{fig:H}, 95\% of the density has $r_s(x) = (2\,n(x))^{-1}$ between 1 and 8.

%-----------------------------
%-----------------------------

Armed with these parametrizations and tools, we are ready to discover 1d electronic structure.

\sec{Results}

DMRG gives us an excellent tool for finding exact answers within a model 1d world.
Our 1d world is designed to mimic qualitatively the 3d world, not match it exactly.
Below we explain some important differences
between our model 1d systems and real 3d systems, starting with the simplest element.

\ssec{One-electron atoms and ions}
As we already mentioned, we find that the energy of the soft-Coulomb hydrogen atom 
is $E(H) = -0.66977714$, accurate to 1 microhartree.
Its ground-state energy is similar to the 3d hydrogen atom energy of $-0.5$~a.u.
Because the potential and wavefunction is much smoother, the kinetic energy is only 0.11 a.u.,
as opposed to 0.5 a.u.\ in 3d.  Since the potential does not scale homogeneously,
the virial theorem in 1d does not yield a simple relation among energy components,
unlike in 3d.

Again because of the lack of simple scaling, hydrogenic energies do not scale
quadratically for our system.   A simple fit of energies for $Z \ge 1$ yields:
\ben
E_Z \approx -Z + \sqrt{Z}/2 - {2}/{9} + {\alpha_1}/{\sqrt{Z}},~~~(N=1)
\een
where $\alpha_1=0.0524$ is chosen to make the result accurate for $Z=1$.
The first two coefficients are exact in the large-$Z$ limit, where the wavefunction
is a Gaussian centered on the nucleus.

%-----------------------------
%-----------------------------

A well-known deficiency of approximate density functionals is their 
{\em self-interaction error}.  Because $E\x$ is approximated, usually in some
local or semilocal form, it fails to cancel the Hartree energy for all
one-electron systems.
Thus, within LSD, the electron incorrectly repels itself.
This error can be quantified by looking at how close $E\x\LSD$ 
is to the true $E\x$.
As can be seen in \Tabref{t:1e}, $E\x\LSD$ is about 10\% too small.  For hydrogen, the self-interaction
error is about 30 millihartrees.  By adding in correlation, this error is slightly reduced, but remains finite.
This is an example of the typical cancellation of errors
between exchange and correlation in LSD.

\begin{table}
%\raggedright
\begin{tabular*}{\columnwidth}{@{\extracolsep{8pt}}ccccccc}
%|c|c|cc|ccc|}
%@{\extracolsep{\fill}}lll
\hline
system&~~$T$~~&$E$&\!\!$E\LSD$&~~$E\x$~&~$E\LSD\x$~&$~E\c\LSD$~~\\
\hline
H	& 0.111 & -0.670 &	-0.647		&	-0.346	& 	-0.311	&	-0.007\\
He$^+$& 0.192 & -1.483 &	-1.455	&		-0.380	&	-0.343	&	-0.006\\
Li$^{++}$	&0.258 & -2.336 &-2.304	&	-0.397	&	-0.359	&	-0.005\\
Be$^{3+}$& 0.316 & -3.209 &-3.176& 	-0.408	&	-0.369	&	-0.005\\
\hline
\end{tabular*}
\caption{
Exact and LSD results for 1d one-electron systems.
\label{t:1e}
}
\end{table}

As a result of self-interaction error, the LSD electron density spreads out too much, as shown in \Figref{fig:H}.  
In this figure we can also see how the LSD KS potential fails to replicate the true KS potential,
which for hydrogen is the same as the external $v(x)$.  And although the LSD KS potential is almost
parallel to $v(x)$ where there is a large amount of density, it decays too rapidly as $|x|\to\infty$.
What this adds up to, both in 1d and in 3d, is that LSD will not bind another electron easily, if at all.
We will return to this point when considering anions.

\ssec{Two-electron atoms and ions}

For two or more electrons, the HF approximation is not exact.  
The traditional quantum chemistry definition of correlation is the error made by HF:
\ben
E\c\QC = E - E\HF.
\een
In \Tabref{t:2e}, we give accurate energy components for two-electron systems;
recall that the components do not satisfy a virial theorem in our 1d systems.
The total energy can be fit just as for one-electron systems, but now:
\ben
E_Z \approx -2\, Z + \sqrt{Z} +c_0 -{\alpha_2}/{\sqrt{Z}},~~~(N=2) %keeping sqrt{Z} coefficient = 1
\een
where $c_0 = 0.507$ and $\alpha_2 = 0.235$.  The HF energies may be fit 
with $c_0\HF = 0.476$ and $\alpha_2\HF = 0.167$.  These fits are not accurate enough
to give the large $Z$ behavior of $E\c\QC$, which seems to vanish as $Z\to\infty$.
For 3d two-electron systems, the correlation
energy scales to a constant at large $Z$ \cite{FTB00}.
Overall, $|E\c\QC|$ is much smaller in 1d than in 3d. 
Rather than the dimensionality, it is the soft nature of our Coulomb
interactions that
causes the reduction in correlation energy compared to 3d. The exact
wavefunctions in
3d have cusps whenever two electrons of opposite spin come together,
caused by the divergence of the electron-electron interaction.  This
cusp-related correlation
is sometimes called dynamic correlation; any other correlation, involving larger
separations of
electrons, is called static \cite{HG11}. (Note that the distinction between static
and dynamic correlation
is not precise.) Our soft-Coulomb potential has no divergence and
induces no cusps, so
dynamical correlation is minimal. There is little static correlation
in tightly bound closed shell
systems, such as our 1d Li$^+$ and Be$^{++}$, so $|E\c\QC| \ll |E|$.
In contrast, for H$^-$, where one electron is loosely bound,
one expects most of the correlation to be static even in 3d,
and one sees large and similar $E\c$ values in 1d and 3d.
In \Secref{s:cor}, we discuss some quantitative measures of strong correlation.

\begin{table}
\begin{tabular*}{\columnwidth}{@{\extracolsep{3pt}}ccccccc}
\hline
system 		&   ~~~~$T$~~~&  ~~~$V$~~~	& ~~~$V\ee$~~~& ~~~~$E$~~~	& ~~$E\HF$~~ & ~~$E\c\QC$~~\\
\hline
	H$^-$	&	0.115	&	-1.326	&	0.481	&	-0.731	&	-0.692 & -0.039	\\
He		&	0.290	&	-3.219	&	0.691	&	-2.238	&	-2.224 & -0.014	\\
Li$^+$		&	0.433	&	-5.084	&	0.755	&	-3.896	&	-3.888 & -0.008\\
Be$^{++}$	&	0.556	&	-6.961	&	0.790	&	-5.615	&	-5.609 &-0.006	\\
\hline
3d H$^-$	&	0.528	&	-1.367	&	0.311	&	-0.528	& -0.488 	& -0.042	\\
3d He		&	2.904	&	-6.753	&	0.946	&	-2.904	&	-2.862	 & -0.042	\\
3d Li$^+$	&	7.280	& -16.13	&     	1.573	&	-7.280	&-7.236		 &-0.043 	\\
3d Be$^{++}$	& 	13.66	&	-29.50	&	2.191	&	-13.66	&-13.61		 & -0.044	\\	
\hline
\end{tabular*}
\caption{
Exact and HF two-electron atoms and ions, in 1- and 3-d (exact data from \Ref{UG94}, Li$^+$
is fit quadratically to surrounding elements, and HF data from \Refs{CCCBDB,DHCU91}).
\label{t:2e}
}
\end{table}

Next we study the {\em exact} Kohn--Sham DFT energy components of these two-electron systems.  
Here we need the DFT definition of correlation, which differs slightly from the traditional quantum
chemistry version:
\bea
E\c &=& E - (T\s + V + U + E\x) \nonumber \\ 
&=& T\c + U\c ,
\label{Ec}
\eea
where $E\x$ is the exchange energy of the exact KS orbitals, $T\s$ is their kinetic energy,
$U$ is the Hartree energy,
$T\c = T - T\s$ is the kinetic correlation energy, and $U\c = V\ee - U  - E\x$ is the potential
correlation energy.  All these functionals are evaluated on the exact ground-state density,
with numerical results found in \Tabref{t:2eDFT}.
The difference between the quantum chemistry $E\c\QC$
and the DFT $E\c$ is never negative and typically much smaller than $|E\c|$ \cite{GPG96}.  For the two-electron
systems of \Tabref{t:2e} and \Tabref{t:2eDFT}, the difference is zero to the given accuracy for
all atoms and ions besides 1d H$^-$.  For our systems, just as in 3d, $E\c\QC-E\c$ vanishes as $Z\to\infty$.
All the large DFT components ($T\s$, $U$, $E\xc$) are typically smaller than their 3d
counterparts and scale much more weakly with $Z$.  However, our numerical results
suggest $T\c \to -E\c$ as $Z\to\infty$, just
as in 3d.

\begin{table}

\begin{tabular*}{\columnwidth}{@{\extracolsep{3pt}}ccccccc}
%|c|ccc|ccc|}

\hline
system 	&  ~~~$T\s$~~~	&  ~~~$U$~~~	&  ~~~$E\xc$~~~&  ~~~$E\x$~~~& ~~~$E\c$~~~&~~~$T\c$~~~ \\

\hline
	H$^-$	&	0.087	&	1.103	&	-0.595	&	-0.552	&	-0.043	&	0.028		\\
He		&	0.277	&	1.436	&	-0.733	&	-0.718	&	-0.014	&	0.013		\\
Li$^+$		&	0.425	&	1.542	&	-0.779	&	-0.771	&	-0.008	&	0.008		\\
Be$^{++}$	&	0.551	&	1.601	&	-0.806	&	-0.801	&	-0.006	&	0.005	\\
\hline
3d H$^-$	&	0.500	&	0.762	&	-0.423	&	-0.381	&	-0.042	&	0.028	\\
3d He		&	2.867	&	2.049	&	-1.067	&	-1.025	&	-0.042	&	0.037	\\
3d Li$^+$ 	& 	7.238  	&	3.313	&	-1.699	&       -1.656	&	-0.043	& 0.041		\\
3d Be$^{++}$ 	&	13.61	&	4.553	&	-2.321	&	-2.277	&	-0.044	& 0.041		\\
\hline
\end{tabular*}
\caption{
Energies of the exact KS system for two-electron atoms and ions.  
3d data (Li$^+$ fitted) from \Ref{UG94}.
\label{t:2eDFT}
}
\end{table}

%-----------------------------
\begin{figure}[htb]
\includegraphics[width=\columnwidth]{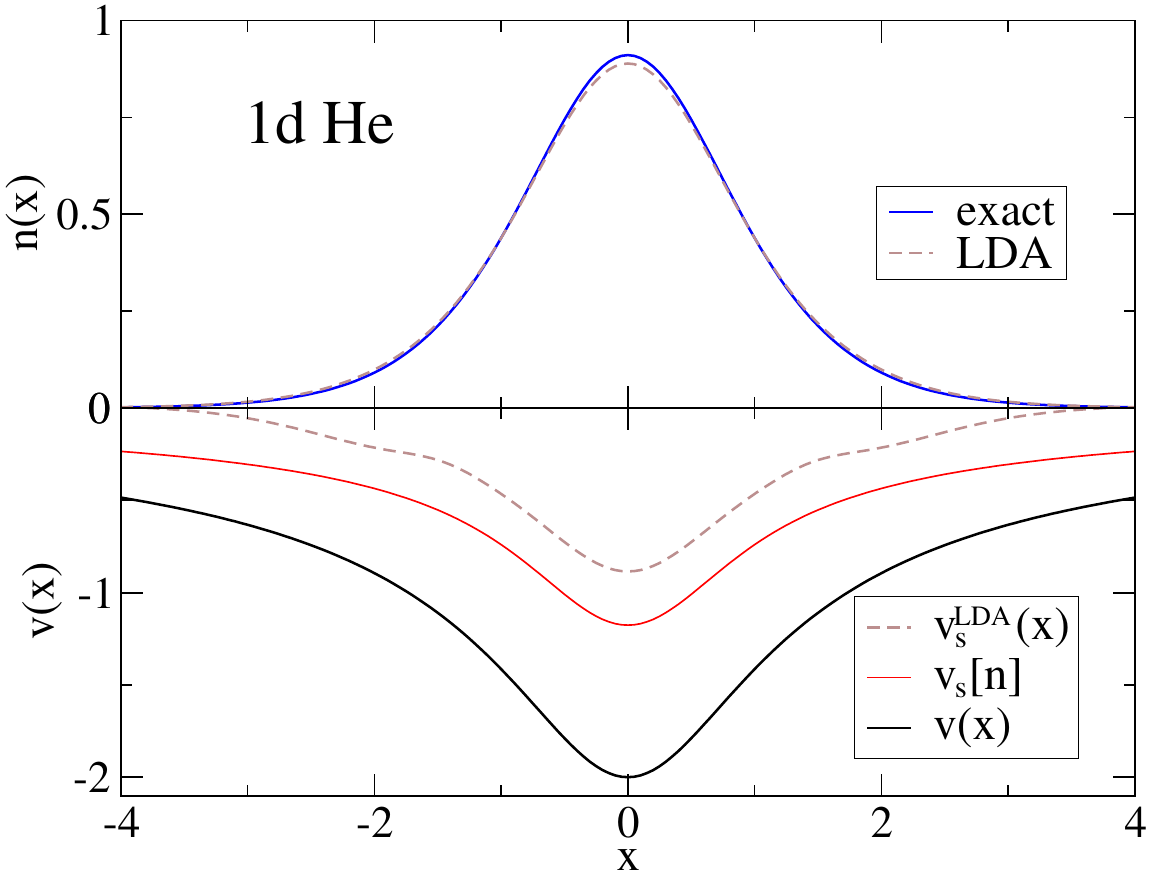} %with KS lda potential and LDA density
\caption{
The exact KS potential for a model helium density found from interacting electrons in 1d,
as well as the LDA density and LDA KS potential found self-consistently.
\label{fig:He}
}
\end{figure}
%-----------------------------

To obtain the KS energies for a given problem, 
we require the KS potential, which is found by inverting the KS equation.
For one- or two-electron systems, this yields:
\ben
v\s(x) = \dfrac{1}{2 \sqrt{n(x)}} \dfrac{d^2 }{dx^2}\sqrt{n(x)}.\quad{(N\le2)} \label{vsN2}
\een
For illustration, consider the exact KS potential of 1d helium in \Figref{fig:He}.
Inverting a density to find the KS potential has also been done for small systems in
3d, where QMC results for a correlated electron density have proven extremely useful \cite{UG94}.
One can find simple and useful constraints on the KS potential by studying 
the large and small $r$ behavior of the exact result \cite{UG94,LPS84}.
In 3d, for large $r$, the Hartree potential screens the nuclear potential,
and the exchange-correlation potential goes like $-1/r$ \cite{AP84}.
In 1d, the softness of the Coulomb potential is irrelevant, so the
Hartree potential screens the nuclear potential for large $|x|$ as in 3d.  Though it
seems likely, we have no proof that the exchange-correlation potential for the soft-Coulomb
interaction should tend to $-1/|x|$ for large $|x|$, analogous to the 3d Coulomb result.  
To check this would require extreme numerical precision in the density far from the atom, due to the need
to evaluate \Eqref{vsN2} where the density is exponentially small.
Instead, in Figures \ref{fig:H2} and \ref{fig:He}, we require the KS potential to go as $-b/|x|$
once the density becomes too small (around $n \approx 10^{-5}$, which happens at $|x| \approx 6$
for helium), 
and we choose $b$ to enforce Koopmans' theorem for KS-DFT.
The actual value of $b$ has no visible effect on the density on the scale of these figures.

We now consider the performance of LDA for these two-electron systems, starting
with how well LDA replicates the true KS potential.
Though the LDA density is only slightly different from the exact density on the scale of \Figref{fig:He},
the LDA potential clearly decays too rapidly (exponentially) at large $r$ and
is too shallow overall, just as in 3d \cite{UG94}.  Like the hydrogen atom discussed earlier,
this is a result of self-interaction error.
LDA energy results are given in \Tabref{t:2eLDA}.  Clearly LDA becomes relatively
more accurate as $Z$ grows, because XC becomes an ever smaller fraction of the total energy.
Comparing Tables \ref{t:2eDFT} and \ref{t:2eLDA}, we also see that LDA underestimates the true 
X contribution by about 10\%, while overestimating
the correlation contribution, so that XC itself has lower error than either, i.e., a
cancellation of errors.

\begin{table}[b]
%\begin{tabular}{|c|cc|ccc|}
\begin{tabular*}{\columnwidth}{@{\extracolsep{4pt}}cccccc}

\hline
 system & ~~$E\LDA$~~	&	\% error	& ~~$E\xc\LDA$~~	&~~$E\x\LDA$~~	& ~~$E\c\LDA$~~\\

\hline
H$^-$	&	-0.708  &	-3.1\%	&	-0.601	&	-0.536	&	-0.065	\\
He	&	-2.201	&	-1.7\%	&	-0.690	&	-0.646	&	-0.044	\\
Li$^+$	&	-3.850	&	-1.2\%	&	-0.731	&	-0.696	&	-0.035	\\
Be$^{++}$&	-5.564	&	-0.9\%	&	-0.753	&	-0.723	&	-0.030	\\	
\hline
3d H$^-$&	-0.511	&	-3.2\%	&	-0.419	&	-0.345	&	-0.074	\\
3d He	&	-2.835	    &	-2.4\%	&  	-0.973	&	-0.862	&	-0.111 	\\
3d Li$^+$&	-7.143	&	-1.9\% &	-1.531	&	-1.396	&	-0.134	\\	
3d Be$^{++}$&	-13.44	&	-1.2\% &	-2.082	&	-1.931	&	-0.150	\\
\hline
\end{tabular*}

\caption{
LDA for 2 electron systems.  H$^-$ does not converge in 1d or 3d; the results are taken
from using the LDA functional on the HF density \cite{LB10}.
3d LDA data from Engel's OEP code \cite{ED99}.
\label{t:2eLDA}
}
\end{table}

%-----------------------------
\begin{figure}[tb]
\includegraphics[width=\columnwidth]{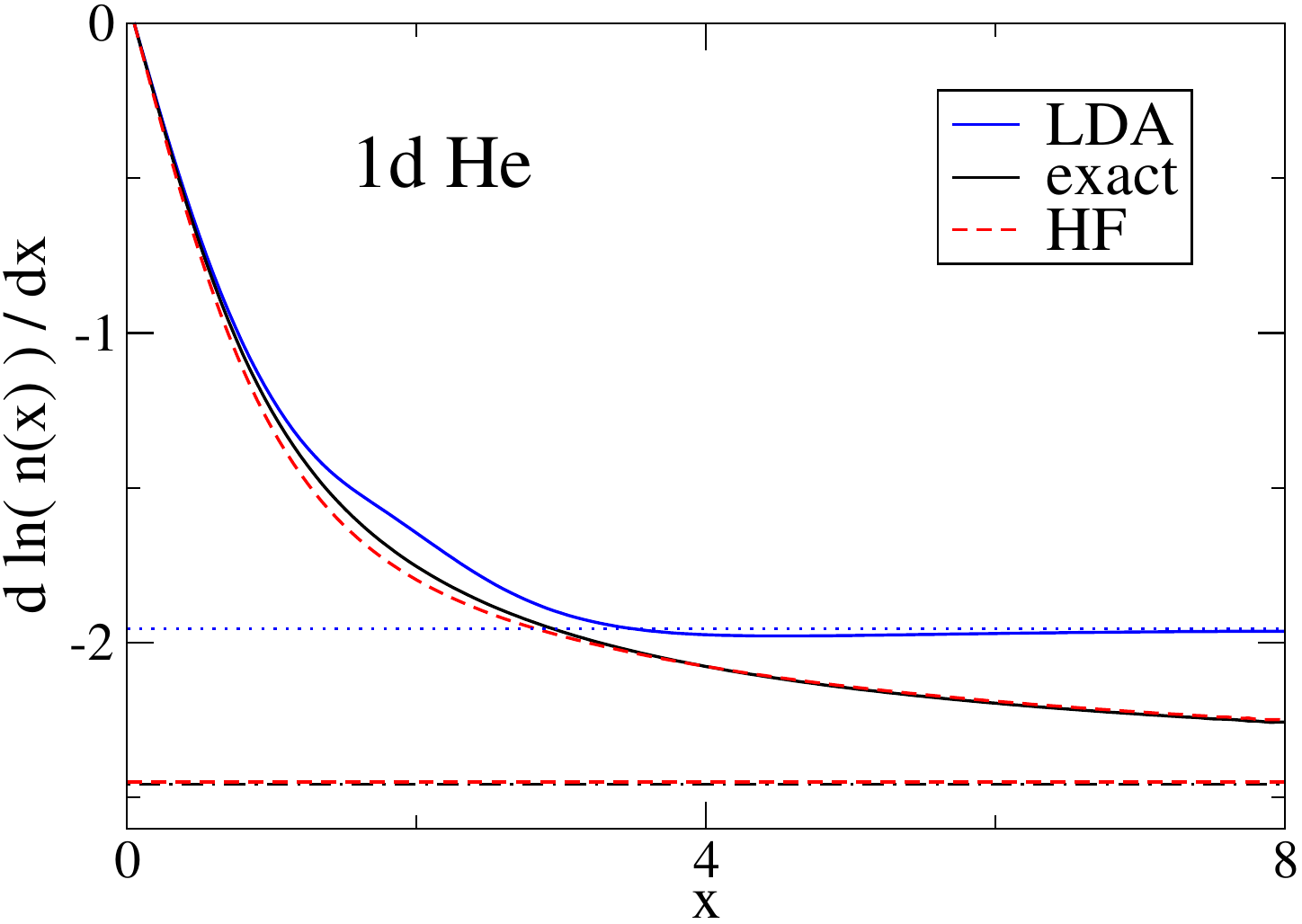}
\caption{The differential logarithmic decay of the helium atom density for various methods.  
The horizontal, dashed lines correspond to the asymptotic decay constants.
}
\label{fig:HeDLOGnDx}
\end{figure}
%-----------------------------

Much insight into density functionals has been gained by studying the asymptotic
decay of densities and potentials far from the nucleus \cite{LPS84}.
In \Figref{fig:HeDLOGnDx}, we plot $dn/dx/n$ to emphasize
the asymptotic decay of the exact, LDA, and HF helium densities.
The HF density is very accurate compared to the LDA density.  For large $x$, the HF density has very nearly the same
behavior as the exact density, and both approach their asymptotes very slowly.  By contrast, the LDA density
reaches its asymptote by $x \approx 4$.
For each approximate calculation, its asymptotic decay constant $\gamma$ can be found using the
highest occupied molecular orbital (HOMO) energy:  $\gamma = -2\sqrt{-2 \epsilon_\text{HOMO}}$.  The asymptote for the
exact curve can be found using the ionization potential $I = E(N-1) - E(N)$ of the system, which determines
the density decay:  $\gamma = -2\sqrt{2I}$.  Because the HF asymptote lies nearly on top of the exact asymptote,
Koopmans' theorem---or $I \cong -\epsilon_\text{HOMO}^\text{HF}$---is extremely accurate for 1d helium.
We list both HOMO and total energy differences in \Tabref{t:2eIeHOMO}.

\begin{table}

%\begin{tabular}{|c|cc|cc|c|}
\begin{tabular*}{\columnwidth}{@{\extracolsep{\fill}}cccccc}
\cline{2-6}
\multicolumn{1}{c}{} & \multicolumn{2}{c}{LSD}	&	\multicolumn{2}{c}{HF}	&	exact	\\
\hline
system	&~-$\epsilon_\text{HOMO}$    &~~~$I$~~~~& 
                    ~-$\epsilon_\text{HOMO}$  &  ~~~$I$~~~~ 
                                        &$I=$-$\epsilon_\text{HOMO}$ \\
\hline
H$^-$    &	|	    &	0.062   &   0.054  &    0.022   &	0.061	\\
He	    &	0.478	&	0.747   &   0.750   &   0.741   &	0.755\\
Li$^+$	&	1.242	&	1.546   &   1.557  &    1.552   &   1.560	\\
Be$^{++}$ &	2.064	&	2.389	&  2.404   &	2.400	&   2.406	\\
\hline
\end{tabular*}
\caption{
1d HOMO eigenvalues and ionization potentials of two-electron atoms and ions, for the exact functional, LDA, and HF.
LDA does not converge for H$^-$ anion, but LDA energies can be found using the HF density \cite{LB10}.
\label{t:2eIeHOMO}
}
\end{table}

There is a long history of studying two-electron ions in DFT, including the smallest anion H$^-$,
which presents interesting conundrums for approximate functionals \cite{SRZ77,KSB11}.
Looking at the ionization energies of these 2 electron systems, we can extrapolate the critical
nuclear charge necessary for binding two electrons, i.e., figuring out the $Z$ value for which $I=0$.  
This happens around $Z = 0.90$ in 1d, and around $Z=0.91$ in 3d \cite{BJ03}.
Within LDA, the critical value is above $Z=1$, because H$^-$ will not
bind.  DFT approximations have a hard time binding anions---both in 1d and in 3d---due to self-interaction error.
A way to circumvent this problem is to take the HF anion, which binds an extra electron,
and evaluate the LDA functional on its density.  As seen in \Tabref{t:2eIeHOMO}, this approach is far better
than either taking total energy differences or the negative of the HOMO energy from HF
alone, just as in 3d \cite{LB10}.
As in 3d, $-\epsilon_\text{HOMO}\LSD$ is useless as an approximation to $I$.
The HF results $-\epsilon_\text{HOMO}\HF$ and $I\HF$ are close to each other and closest to $I$ for larger $Z$;
but $I\LSD$ does very well for small $Z$, and is best for H$^-$.

\ssec{Many-electron atoms}

\begin{figure}[t]
\includegraphics[width=\columnwidth]{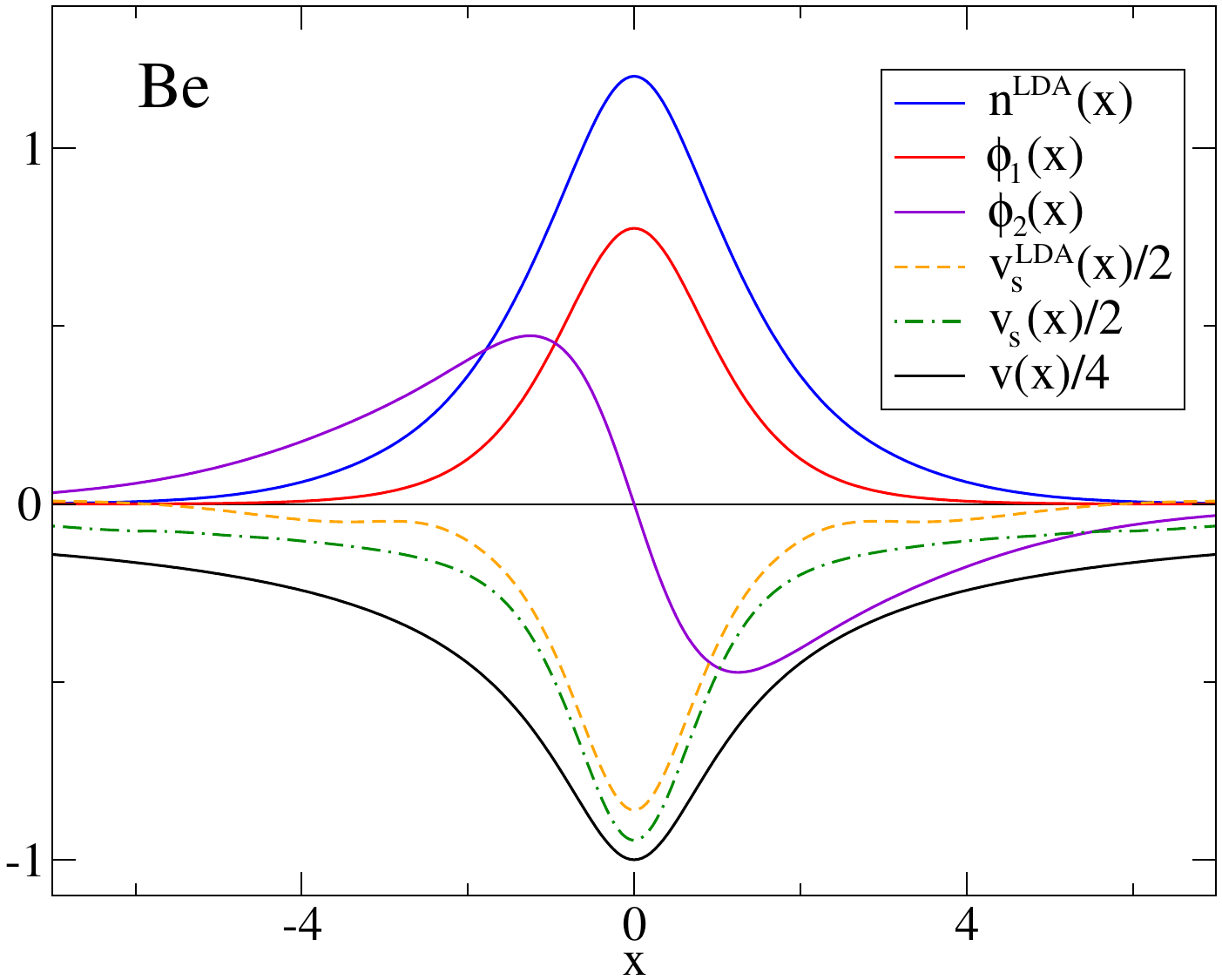}
\caption{An LDA ``beryllium'' atom, complete with the LDA orbitals $\phi_j(x)$, LDA KS potential $v\s\LDA(x)$, exact KS potential
$v\s(x)$, and external potential $v(x)$.
The density $n\LDA(x)$ was found self-consistently using the LDA method, and barely differs from the true $n(x)$ on this scale.
}
\label{fig:BeLDA}
\end{figure}

Before looking at larger atoms, a word of caution.
In 3d systems, degeneracies in orbitals with different angular momentum quantum numbers
produce interesting shell structure.  In 1d, there is no angular momentum---each 1d
shell is either half-filled or filled---so it is not
clear which real elements our model 1d atoms correspond to.  The first
three 1d elements might well be called hydrogen, helium, and lithium; but the fourth 1d
element may behave more like neon than beryllium.  To be consistent with \Ref{HFCV11},
we call it beryllium.  To showcase
1d Be, consider its LDA treatment in \Figref{fig:BeLDA}.  
The exact KS potential is also plotted, and the LDA KS potential roughly differs 
only by a constant in the high density region, just as with hydrogen (\Figref{fig:H}) and helium (\Figref{fig:He}).
In the low density
regions, the LDA correlation potential is the dominant piece of the LDA KS potential.

%-----------------------------
%-----------------------------

As we increase the number $N$ of electrons in our systems, correlation also increases, but HF
theory is still better than LDA until $N=4$.  
Exact and HF data for many-electron atoms can be found in \Tabref{t:me}, and
LDA data in \Tabref{t:meLDA}.
Despite good agreement with all other data, we did not find He$^-$ or Li$^-$ to bind as in \Ref{HFCV11},
neither in HF nor DMRG, nor LSD (not surprisingly).  When using energy differences to calculate the ionization
energy, HF outperforms LSD until beryllium, as can be seen in \Tabref{t:meIeHOMO}.  For these
systems, $I\LSD > I > I\HF$:  LSD overestimates the ionization energy, and HF underestimates it---just as in 3d.  As with
the fewer electron systems,
the LSD HOMO energies are not a good way to estimate $I$, whereas the HF HOMO energies are.

\begin{table}
%\begin{tabular}{|c|ccc|ccc|}
\begin{tabular*}{\columnwidth}{@{\extracolsep{4pt}}ccccccc}
\hline
system 		&   ~~~~$T$~~~&  ~~~$V$~~~	& ~~~$V\ee$~~~& ~~~~$E$~~~	& ~~$E\HF$~~ & ~~$E\c\QC$~~\\
\hline
	Li	&	0.625	&	-6.484	&	1.648	&	-4.211	&	-4.196  & -0.015	\\
	Be$^+$	&	0.922	&	-9.240	&	1.864	&	-6.454	&	-6.445	& -0.010	\\
	Be	&	1.127	&	-11.13	&	3.219	&	-6.785	&	-6.740	& -0.046	\\
\hline
\end{tabular*}
\caption{
Exact and HF many-electron atoms and ions, in 1d.
\label{t:me}
}
\end{table}

\begin{table}[htb]
%\begin{tabular}{|c|cc|ccc|}
\begin{tabular*}{\columnwidth}{@{\extracolsep{\fill}}cccccc}

\hline
 & ~~$E\LSD$~~	&	\% error	& ~~$E\xc\LSD$~~	&~~$E\x\LSD$~~	& ~~$E\c\LSD$~~\\

\hline
Li	&	-4.179  &	-0.8\%	&	-1.044	&	-1.004	&	-0.041	\\
Be$^+$	&	-6.410	&	-0.7\%	&	-1.117	&	-1.086	&	-0.031	\\	
Be	&	-6.764	&	-0.3\%	&	-1.450	&	-1.376	&	-0.075	\\
\hline

\end{tabular*}
\caption{
LSD energies for many-electron 1d systems.
\label{t:meLDA}
}
\end{table}

\begin{table}[htb]
%\begin{tabular}{|c|cc|cc|c|}
\begin{tabular*}{\columnwidth}{@{\extracolsep{\fill}}cccccc}
\cline{2-6}
\multicolumn{1}{c}{} & \multicolumn{2}{c}{LSD}	&	\multicolumn{2}{c}{HF}	&	exact	\\
\hline
system	&~-$\epsilon_\text{HOMO}$    &~~~$I$~~~~& 
                    ~-$\epsilon_\text{HOMO}$  &  ~~~$I$~~~~ 
                                        &$I=$-$\epsilon_\text{HOMO}$ \\
\hline
Li    &	0.166   &	0.329   &   0.316  &    0.308   &	0.315	\\
Be$^+$& 0.628	&	0.846	&   0.842  &    0.835	&	0.839	\\
Be    & 0.162	&	0.353	&   0.313  &	0.295	&	0.331	\\

\hline
\end{tabular*}
\caption{
Many-electron ionization energies for LSD, HF, and exact 1d systems.
\label{t:meIeHOMO}
}
\end{table}

To find the KS energy components for these many-electron ($N > 2$) systems, we again require the exact KS potential.
For these systems, \Eqref{vsN2} is no longer valid, so we must find the KS potential another way.  
The simplest procedure is to use guess-and-check,
adjusting the KS potential until its density can no longer be distinguished from the target density found using DMRG.
Updates to the KS potential can be more or less sophisticated without changing the final result in the region
where the density is large; in the low-density region, however, two very different KS potentials can give rise to
densities that are indistinguishable on the scale of our figures.  However, the KS energy components do not rely significantly
on the behavior of the KS potential out in the low-density region.  
In \Tabref{t:meDFT}, the exact KS energies for
some many-electron systems are tabulated.  For Li and Be$^+$, spin-DFT is used, but the spin-dependent
energy components (such as $T\s^\sigma$) are summed together to give a spin-independent energy.

\begin{table}

%\begin{tabular}{|c|ccc|ccc|}
\begin{tabular*}{\columnwidth}{@{\extracolsep{\fill}}ccccccc}

\hline
 	&  ~~~$T\s$~~~	&  ~~~$U$~~~	&  ~~~$E\xc$~~~&  ~~~$E\x$~~~& ~~~$E\c$~~~&~~~$T\c$~~~ \\

\hline
	Li	&	0.611	&	2.749	&	-1.087	&	-1.071	&	-0.016	&	0.014		\\
	Be$^+$	&	0.912	&	3.042	&	-1.168	&	-1.157	&	-0.011	&	0.009		\\
	Be	&	1.091	&	4.736	&	-1.481	&	-1.430	&	-0.051	&	0.036		\\
\hline

\end{tabular*}
\caption{
Energies of the exact KS system for many-electron 1d atoms and ions.  
\label{t:meDFT}
}
\end{table}

\begin{figure}[t]
\includegraphics[width=\columnwidth]{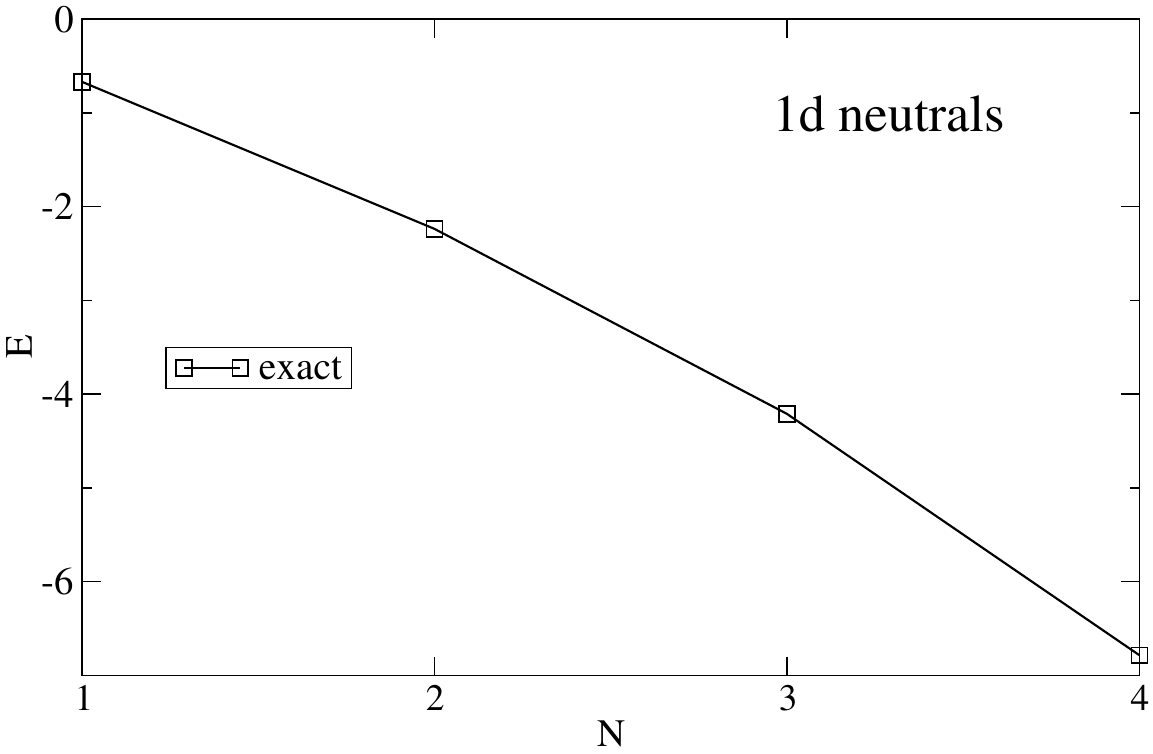}
\caption{Energies of neutral atoms in 1d.
\label{fig:largeZ}
}
\end{figure}

The study of the energies of neutral atoms as $N = Z\to \infty$ is important due to the
semiclassical result being exact in that limit \cite{ELCB08}.  In this limit, the oldest
of all density functional approximations, Thomas--Fermi (TF) theory, becomes exact \cite{L81}.  However, due to a lack of 
scaling within the soft-Coulomb interaction, the large $Z$ limit of the energy is non-trivial, 
making a semiclassical treatment difficult.
A plot of the neutral atom energies as a function of $N$ appears in \Figref{fig:largeZ}.
On this scale, both the LDA and HF results lie nearly on top of the exact curve.

%-----------------------------
%-----------------------------

\ssec{Equilibrium properties of small molecules}\label{s:smallmole}

\begin{figure}[t]
\vspace{0.5cm}
\includegraphics[width=\columnwidth]{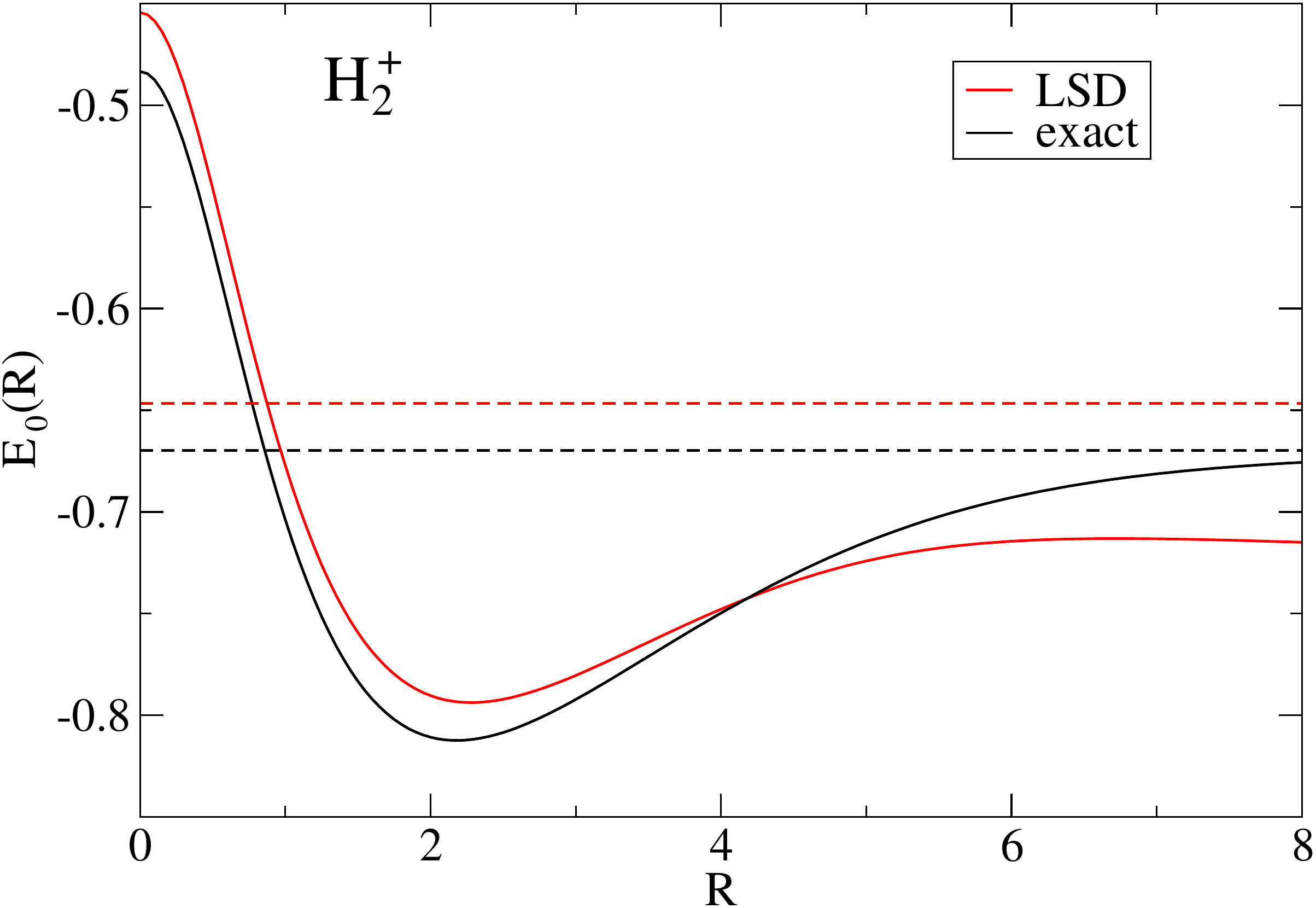}
\caption{The binding energy curve for our 1d model H$_2^+$, shown with an absolute energy scale, and 
with nuclear separation $R$; horizontal dashed lines indicate the energy of a single H atom.}
\label{fig:H2pBindingEnergy}
\end{figure}

We now briefly discuss small molecules near their equilibrium separation.
In order to find the equilibrium
bond length for our 1d systems, we take the nuclei to be interacting via the soft-Coulomb interaction, just like the electrons.
Given this interaction, consider
the simplest of all molecules: the H$_2^+$ cation.  HF yields the exact
answer, and LSD suffers from self-interaction (more generally, a delocalization error \cite{CMY08}).
A plot of the binding energy is found
in \Figref{fig:H2pBindingEnergy}.  Because the nuclear-nuclear repulsion is softened, the binding energy
does not diverge as the internuclear separation $R$ goes to zero.
As seen in \Tabref{t:H2nums}, LSD overbinds slightly and produces bonds that are too long between H atoms,
which is also the case in 3d.
The curvature of the LSD binding energy is too small near equilibrium,
which makes for inaccurate
vibrational energies, especially in 3d.  This can also be seen in \Tabref{t:H2nums}.
Finally, we note that the energy of stretched H$_2^+$ does not tend to that of 
H within LSD, due to delocalization error  \cite{CMY08}.

%-----------------------------
%-----------------------------

\begin{table}

%\begin{tabular}{|c|c|c|c|}
\begin{tabular*}{\columnwidth}{@{\extracolsep{\fill}}cccc}
\cline{2-4}

\multicolumn{1}{c}{}	& ~~~HF~~~	& ~~LSD~~	&	~~exact~~	\\
\hline
~~system~~	& \multicolumn{3}{c}{$D\e$ (eV)} \\
\hline
H$_2^+$	&	3.88  ~(0\%)	&  4.00	 ~(3\%)	&	3.88	\\
3d H$_2^+$&	2.77 ~(0\%)	& 2.89 ~(4\%)	&	2.77	\\
\hline
H$_2$	&	2.36  (-23\%)	&  3.53	 (15\%)	&	3.07	\\
3d H$_2$&	3.54 (-25\%)	& 4.80 ~(1\%)	&	4.75	\\
\hline

system	& \multicolumn{3}{c}{$R_0$} \\
\hline
H$_2^+$	&	2.18  ~(0\%)	&  2.28	 ~(4\%)	&	2.18	\\
3d H$_2^+$&	2.00 ~(0\%)	& 2.18 ~(9\%)	&	2.00	\\
\hline
H$_2$	&	1.50 ~(-6\%)	& 1.63 ~(2\%)  	&	1.60	\\
3d H$_2$&	1.41 ~(1\%)	&  1.47 ~(5\%)	&	1.40	\\

\hline
system	& \multicolumn{3}{c}{$\omega$ ($\times10^3$ cm$^{-1}$)} \\
\hline
H$_2^+$	&	2.2 ~(0\%)	&  2.0 ~(-9\%)	&	2.2	\\
3d H$_2^+$&	2.4 ~(0\%)	&  1.9 (-21\%)	&	2.4	\\
\hline
H$_2$	&	3.3 ~(6\%)	&  3.0 ~(-3\%)  &	3.1	\\
3d H$_2$&	4.6 ~(5\%)	&  4.2 ~(-5\%)	&	4.4	\\

\hline
\end{tabular*}

\caption{
Electronic well depth $D\e$, equilibrium bond radius $R_0$, and 
vibrational frequency $\omega$  for the H$_2^+$ and H$_2$ molecules, 
with percentage error in parentheses.
Exact 3d H$_2$ results taken from \Ref{KW68};
the remaining 3d values are from \Ref{CCCBDB} using the aug-cc-pVDZ basis set \cite{aug-cc-pVXZ}.
\label{t:H2nums}
}
\end{table}

\begin{figure}[t]
\includegraphics[width=\columnwidth]{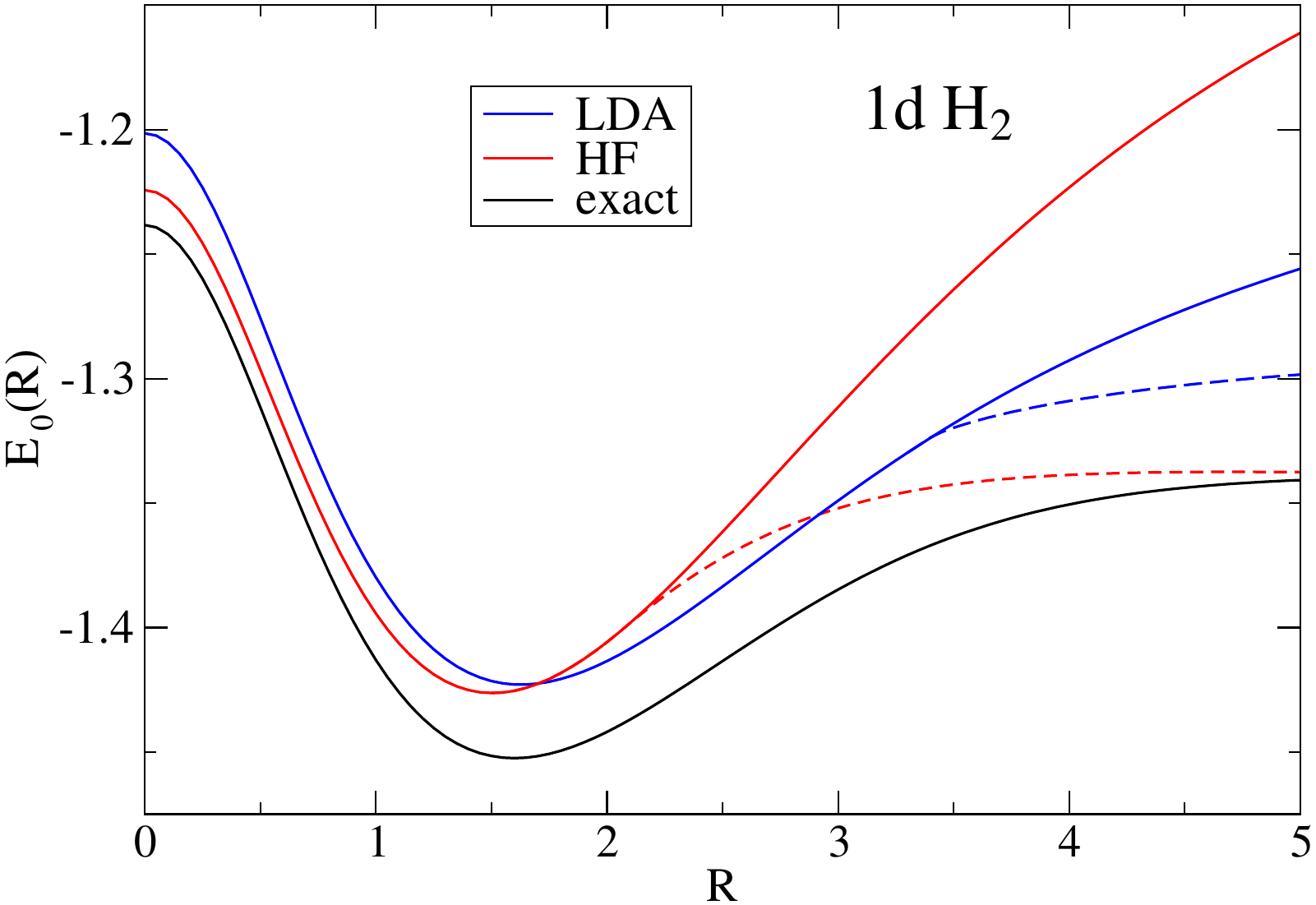}
\caption{The binding energy curve for our 1d model H$_2$, shown on an absolute energy scale, 
with nuclear separation $R$.  Dashed curves represent unrestricted calculations.
}
\label{fig:H2BindingEnergy}
\end{figure}

Next we consider H$_2$.
A plot of the binding energy  is found in \Figref{fig:H2BindingEnergy}; the large
$R$ behavior will be discussed in the following section.
Just as in 3d, HF underbinds while LDA overbinds; HF bonds are too short, and LDA bonds are too long.
Further, HF yields vibrational frequencies which are too high, and LDA are a little small, which
is the case both in 1d and 3d.
All of these properties can be seen in \Tabref{t:H2nums}.

%-----------------------------
%-----------------------------

\ssec{Quantifying correlation}\label{s:cor}

It is often said that DFT works well for weakly correlated systems, but fails when correlation is too strong.
Strong static correlation, which occurs when molecules are pulled apart, is also identified with strong correlation
in solids \cite{CMY08}.  Functionals that can accurately deal with strong static correlation in stretched molecules can also accurately yield the band gap of a solid \cite{SCY08,ZCMP11}.
Most DFT methods, however, fail in these situations. 
To see these effects in 1d, we shall now examine three descriptors
of strong correlation, which will be 0 when no correlation is present and close to 1 when strong correlation is present.

A simple descriptor of strong correlation is simply to calculate the ratio
of correlation to exchange: 
\ben
\alpha = \dfrac{E\c}{E\x}.\label{alpha}
\een 
In the limit of weak electron-electron repulsion, $\alpha$ goes to zero for closed-shell systems, and HF becomes exact.  For example,
for the two-electron atoms and ions in \Tabref{t:2eDFT}, $\alpha$ goes to zero as $Z$ increases.
In \Tabref{t:H2cor}, we compute $\alpha$ for various bond lengths of the hydrogen molecule, both in 1d and in 3d.
At the equilibrium bond length, $\alpha$ is small, indicating that the HF solution is
very close to the exact.  When the bond is stretched to $R=5$, $\alpha$ increases ten-fold:
a standard HF solution for a bond length of $R = 5$ does not do well at all.
The 1d and 3d results are remarkably similar.
We can also compute $\alpha$ using the LDA functionals for $E\c$ and $E\x$
evaluated on the LDA density; however, $\alpha\LDA$ is not
as good of an indicator for strong correlation as the true $\alpha$ is.

The second descriptor of strong correlation requires first understanding where correlation comes from.
From \Eqref{Ec}, correlation can be separated into two pieces: (1) the kinetic correlation energy $T\c = T - T\s$, due to the small difference between the true kinetic
energy and the KS kinetic energy, and (2) potential correlation energy, $U\c = V\ee - U - E\x$. 
In the limit
of weak correlation in 3d, $U\c\rightarrow -2T\c$, so the ratio:
\ben
\beta = \frac{E\c+T\c}{E\c} \label{beta}
\een
has always been found to be positive, and vanishes in the weakly correlated limit \cite{BPEb97}, which
we have also observed in 1d.  
But if $T\c \ll |E\c|$, we have correlation without the usual kinetic contribution,
which occurs when systems have strong {\em static} correlation.  
For example, in the infinitely stretched limit of
H$_2$, $T\s \to T$ while $E\c$ remains finite, so $\beta \to 1$.
In \Tabref{t:H2cor}, we see that $\beta$ increases as we stretch the H$_2$ molecule,
both in 1d and 3d.
Thus $\beta$ is a natural measure of static correlation in chemistry.

There is another test for strong static correlation, which we can use
on closed-shell systems:  whether an approximate calculation prefers to break
spin-symmetry or not.
This well-known phenomenon occurs when a molecule, such as H$_2$, is stretched
beyond the Coulson--Fischer point \cite{CF49}, and indicates the preference for 
electrons to localize on different atoms.
Spin-symmetry-breaking can be observed in \Figref{fig:H2BindingEnergy},
where the solid (dashed) curves represent restricted (unrestricted) calculations for a stretched hydrogen molecule.
Unrestricted HF/LSD breaks spin symmetry around \mbox{$R=2.1$/3.4},
beyond which the unrestricted solution gives accurate total energies, but very incorrect
spin densities \cite{PSB95}.
The numbers are similar in 3d ($R=2.3$/3.3) \cite{BAb96}.

\begin{table}
%\begin{tabular}{|c|c|ccc|cc|}
\begin{tabular*}{\columnwidth}{@{\extracolsep{\fill}}ccccccc}
\cline{3-7}
\multicolumn{2}{c}{}  &   \multicolumn{3}{c}{1d} &\multicolumn{2}{c}{3d}  \\
\cline{2-7}
\multicolumn{1}{c}{}	& ~~$R$~~      & ~~1.6~~  &~~3.4~~    &   ~~5.0~~     &  ~~ 1.4~~  & ~~  5.0~~  \\
        \hline
~exact~		& ~~$\alpha$~~ &   0.04   &	0.21  &   0.46    & 	 0.06 	&  0.45   \\
			&  ~~$\beta$~~ &   0.21   &  	0.58  & 0.87    &  	0.18	 &  0.89 \\
\hline
LDA			& $\alpha$	&  0.09   &	0.16  &   0.21  & \multicolumn{2}{c}{} \\
\cline{1-5}
\end{tabular*}
\caption{Table of correlation descriptors $\alpha$ and $\beta$ (Eqs.~\eqref{alpha} and \eqref{beta}) for H$_2$ at an equilibrium
and a stretched bond length $R$.  3d data from \Ref{FNGB05}.
\label{t:H2cor}
}
\end{table}

%%%%%%%%%%%%%%%%%%%%%%%%%%%%%%%%%%%%%%%%%%%%%%%%%%%%%%%%%%%%%%%%%%%%%%%%%%%%%%%%%%%%%%%%%%%%%%%%%%%%%%%%%%%%%%%%%%%%%%%%
%%%%%%%%%%%%%%%%%%%%%%%%%%%%%%%%%%%%%%%%%%%%%%%%%%%%%%%%%%%%%%%%%%%%%%%%%%%%%%%%%%%%%%%%%%%%%%%%%%%%%%%%%%%%%%%%%%%%%%%%
%%%%%%%%%%%%%%%%%%%%%%%%%%%%%%%%%%%%%%%%%%%%%%%%%%%%%%%%%%%%%%%%%%%%%%%%%%%%%%%%%%%%%%%%%%%%%%%%%%%%%%%%%%%%%%%%%%%%%%%%

%%%%%%%%%%%%%%%%%%%%%%%%%%%%%%%%%%%%%%%%%%%%%%%%%%%%%%%%%%%%%%%%%%%%%%%%%%%%%%%%%%%%%%%%%%%%%%%%%%%%%%%%%%%%%%%%%%%%%%%%

\sec{Conclusion}

In this paper, we have surveyed basic features of the electronic structure
of a one-dimensional world of electrons and protons interacting via
a soft-Coulomb interaction.  
We have established many key reference values
for future use in calculations of atoms, molecules, and even solids.
This 1d world forms a virtual laboratory for understanding and improving
electronic structure methods.  
A major advantage of the 1d world is provided by DMRG which is extremely
efficient and accurate for such systems, making large system sizes readily 
accessible.  Furthermore, the thermodynamic limit is far more quickly
approached in 1d than in 3d.

But none of this would be useful if, in this 1d world, both exact and approximate
calculations did not behave qualitatively similarly to their 3d analogs.  If bonds
are not formed or ions do not exist, there would be no 1d analog of many of the 
energy differences that are used as real electronic stucture benchmarks.
This work contains an extremely detailed study of the qualitative similarities,
and differences, between this 1d world and our own.

For atoms and cations, we find trends in the exact numbers quite similar to real atoms.
However, densities are more diffuse and correlation is weaker, so that Hartree-Fock
is more accurate than in 3d.  An important
technical difference is that the interaction does not scale simply under coordinate
scaling, so that even hydrogenic atoms do not scale simply with $Z$, and there
is no simple virial relation among the energy components of atoms and ions.
Perhaps the most important caveat is that, while atoms with more than 2 electrons
exist, there are no orbital shells in 1d, so there is no clear analog to specific
real atoms.  Our 1d periodic table has only two columns.

Equally important, if the 1d world is to be useful in studying DFT, is that
the standard approximations work and fail under the same circumstances as in 3d.
We have shown that spin-polarization effects can be much stronger in the 1d uniform
gas than in 3d, and this has an effect on the local (spin) density approximation.
But LSD behaves very similarly to its 3d analog, not just for energetics, but
also in the poor behavior of the potential and HOMO eigenvalue.  

For the prototype molecules, H$_2$ and H$_2^+$, we find LDA working well at
equilibrium, with errors similar to those in 3d.  As the bonds are stretched
and correlation effects grow,
H$_2^+$ shows the usual self-interaction or delocalization error, while
approximate treatments of H$_2$ break symmetry, just as in 3d.  Remarkably,
simple measures of correlation at both equilibrium and stretched bonds are
quantitatively similar to their 3d counterparts.  

These results suggest that understanding electron correlation in this 1d world will
provide insight into real 3d systems, and illuminate
the challenges to making approximate DFT
work for strongly correlated systems.
Other approximate approaches to correlation, such as 
range-separated hybrids \cite{S96}, dynamical mean-field theory \cite{GKKR96}, and LDA+U \cite{AAL97},
can be tested in the future.  Furthermore, fragmentation schemes such
as partition density functional theory (PDFT) \cite{EBCW10}, can be tested
for fully interacting fragments using the exact exchange-correlation
functional, calculated via DMRG.
We expect many further 1d explorations in the future.

With gratefulness, we acknowledge DOE grant DE-FG02-08ER46496 (KB and LW) 
and  NSF grant DMR-0907500 (ES and SW) for supporting this work.

\bibliography{dmrgdft-long}

\end{document}